\documentclass[10pt,conference]{IEEEtran}
\usepackage{ragged2e}
\usepackage{csquotes}
\usepackage{enumitem}
\usepackage[ruled,linesnumbered]{algorithm2e}
\usepackage{cite}
\usepackage{amsmath,amssymb,amsfonts}
\usepackage{algorithmic}
\usepackage{graphicx}
\usepackage{textcomp}
\usepackage{xcolor}
\def\BibTeX{{\rm B\kern-.05em{\sc i\kern-.025em b}\kern-.08em
    T\kern-.1667em\lower.7ex\hbox{E}\kern-.125emX}}

\newcommand{\ssymbol}[1]{^{\@fnsymbol{#1}}}
\usepackage[export]{adjustbox}
\usepackage[T1]{fontenc}

\DeclareMathOperator*{\argmax}{argmax}

\usepackage{mathtools}
\usepackage{array}
\usepackage{float}
\DeclarePairedDelimiter\abs{\lvert}{\rvert}%
\usepackage[english]{babel}

\begin{document}
\title{An MRL-Based Design Solution for RIS-Assisted MU-MIMO Wireless System under Time-Varying Channels\\
%{\footnotesize \textsuperscript{*}Note: Sub-titles are not captured in Xplore and should not be used}

}
\author{\IEEEauthorblockN{Meng-Qian Alexander Wu$^{\dagger}$,  Tzu-Hsien Sang*, Luisa Schuhmacher$^{\mathsection}$,\\ Ming-Jie Guo*, Khodr Hammoud$^{\mathsection}$, Sofie Pollin$^{\mathsection}$}
\thanks{The work of Luisa Schuhmacher has received funding from the European Union's Horizon 2020 Marie Skłodowska Curie Innovative Training Network Greenedge (GA. No. 953775). The work of Khodr Hammoud was funded by the Flemish FWO SBO S001521N IoBaLeT project (sustainable Internet of Battery-Less Things).}
	\IEEEauthorblockA{$^{\dagger}$Department of Electronic and Electrical Engineering, NYCU, Hsinchu, Taiwan, *Institute of Electronics,\\ NYCU, Hsinchu, Taiwan, $^{\mathsection}$ESAT, KU Leuven, Belgium\\
		Email: kawu.ee08@nycu.edu.tw, tzuhsien54120@nycu.edu.tw}
}
\maketitle
\begin{abstract}
	Utilizing Deep Reinforcement Learning (DRL) for Reconfigurable Intelligent Surface (RIS) assisted wireless communication has been extensively researched. However, existing DRL methods either act as a simple optimizer or only solve problems with concurrent Channel State Information (CSI) represented in the training data set. Consequently, solutions for RIS-assisted wireless communication systems under time-varying environments are relatively unexplored. However, communication problems should be considered with realistic assumptions; for instance, in scenarios where the channel is time-varying, the policy obtained by reinforcement learning should be applicable for situations where CSI is not well represented in the training data set.
	In this paper, we apply Meta-Reinforcement Learning (MRL) to the joint optimization problem of active beamforming at the Base Station (BS) and phase shift at the RIS, motivated by MRL's ability to extend the DRL concept of solving one Markov Decision Problem (MDP) to multiple MDPs. 
	We provide simulation results to compare the average sum rate of the proposed approach with those of selected forerunners in the literature. Our approach improves the sum rate by more than 60\% under time-varying CSI assumption while maintaining the advantages of typical DRL-based solutions.
	Our study's results emphasize the possibility of utilizing MRL-based designs in RIS-assisted wireless communication systems while considering realistic environment assumptions.
\end{abstract}


\section{Introduction}

%\subsection{Backgrounds}

	While the fifth generation (5G) wireless communication has been standardized and commercialized worldwide, the volume of global mobile traffic constantly increases. Research has shown that the traffic volume will reach five zettabytes per month by 2030\cite{union2015imt}. To cope with the increasing traffic demands, researchers have started focusing on sixth-generation (6G) wireless communication networks. Considering sustainability, the concept of "green 6G" has been proposed\cite{huang2019survey}. While extending the massive Multi-Input Multi-Output (MIMO) system is the most straightforward solution considering the high data transmission, it can result in high hardware complexity and energy demand to maintain the communication system. Therefore, researchers focus on creating a Smart Radio Environment (SRE). The SRE aims to recycle existing radio waves to meet high data transmission requirements and promote green networks without generating new signals.
	
	Reconfigurable Intelligent Surface (RIS), also known as Intelligent Reflecting Surface (IRS), is an essential component in creating an SRE. It is capable of reflecting and transmitting mmWave signals, which helps overcome the degrading influence of physical obstacles in the environment. Furthermore, RIS is a cost-effective and energy-efficient solution that supports the aim of green 6G. RIS can easily be integrated into existing wireless network systems, providing the ability to create a direct and clear path in the original propagation space. However, to reflect signals, RIS is required to adjust phase shifts constantly, presenting a challenging research problem.
	
	%\subsection{Related Work}
	
	Significant research on RIS-assisted communication with different optimization goals and usage scenarios has been published in the literature. This research can be broadly classified into two main categories: traditional optimization-based\cite{han2019large, wu2019intelligent, wu2019beamforming, huang2018achievable, fu2021reconfigurable, ma2021joint} and Artificial Intelligence-based (AI)\cite{faisal2021deep, huang2020reconfigurable, huang2021multi, feng2020deep, zhu2022deep, zhu2022drl}.
	
	The first optimization-based algorithm, proposed by Huang et al., considered an outdoor scenario with RIS assisting Multi-User Multiple-Input Single-Output (MU-MISO) downlink communication\cite{huang2018achievable}. Huang's research increased the system sum rate by more than 40\% without additional energy consumption. In parallel with Huang's research, Wu et al. proposed two algorithms for RIS with discrete phase shift \cite{wu2019intelligent} and with continuous phase shift \cite{wu2019beamforming}. However, Wu adopted semidefinite relaxation (SDR), a high-complexity technique, to solve the non-convex quadratically constrained quadratic program (QCQP). Following Wu's research, Ma et al. \cite{ma2021joint} proposed an algorithm with the lowest complexity to the best of our knowledge. Optimization-based algorithms give optimized RIS phase shift and beamformer design for given instant Channel State Information (CSI). However, under time-varying channels, optimization-based algorithms must continually update their solutions and thus can become quite inefficient because of the large computational cost associated with the optimization procedure. High time consumption is still a challenging problem for this type of algorithm in time-varying environments.
	
	The first AI-based algorithm proposed by Huang et al.\cite{huang2021multi} utilized the Deep Deterministic Policy Gradient (DDPG) algorithm for RIS-assisted multi-hop multi-user wireless THz communication to provide both digital and analog beamforming. Considering DDPG easily falls in the optimal local solution, Zhu et al.\cite{zhu2022drl} proposed the first Soft Actor-Critic (SAC) based algorithm to jointly optimize both Base Station-RIS-User (BS) association and passive beamforming. Deep Reinforcement Learning (DRL) based algorithms require significantly less computation time. However, all AI-based algorithms consider only time-invariant CSI, leading to poor performance in non-stationary environments.
	
	%\subsection{Overview}
	
    In this paper, we studied the joint optimization of phase shift and beamforming for RIS-assisted systems in non-stationary environments. The main challenge of non-stationary environments is the CSI changes in every timestamp, resulting in the optimization problem varying on a per-timestamp basis. To address such a complicated problem, we considered the CSIs as a group of Markov Decision Processes (MDPs) with similar structures, which can be treated by Meta-Reinforcement Learning (MRL). The main contributions are organized as follows:
    \begin{enumerate}[label=\arabic*)]
    \item We develop the first AI-based design solution for RIS-assisted systems under time-varying channels. The training procedure in which a large amount of time-varying CSIs are fed into the MRL-based algorithm enables the trained result to identify and exploit the proximity between unseen CSIs in testing with experienced CSIs in training. It is a huge step forward from DRL-based algorithms that can only handle experienced CSIs but may perform poorly even with similar CSIs.
    \item We jointly design the RIS phase shift and the beamformer to maximize the downlink average sum rate under time-varying channels and power constraints at BS.
    \item We demonstrate the outstanding performance of the proposed solution under time-varying channels in terms of average downlink sum rate. The time-varying channels are modeled by autoregressive (AR) models.
    \end{enumerate}
    
    The remainder of this paper is organized as follows. Section II introduces the system model of a downlink RIS-assisted system under the assumption of time-varying CSIs and the problem formulation. Then, a RIS phase shift and beamforming design algorithm based on MRL is proposed in Section III with the training-testing procedure and pseudo-code. Section IV shows the simulation results, demonstrating superior performance compared to existing methods. Finally, Section V ends the paper with a short conclusion.

\section{System Model and Problem Formulation}

	In this section, the wireless system model is introduced to develop an energy-efficient solution for digital beamforming and passive phase shift for RIS-assisted communication systems.               
	
\subsection{System Model}
\begin{figure}
		\centering
		\includegraphics[width=0.7\linewidth]{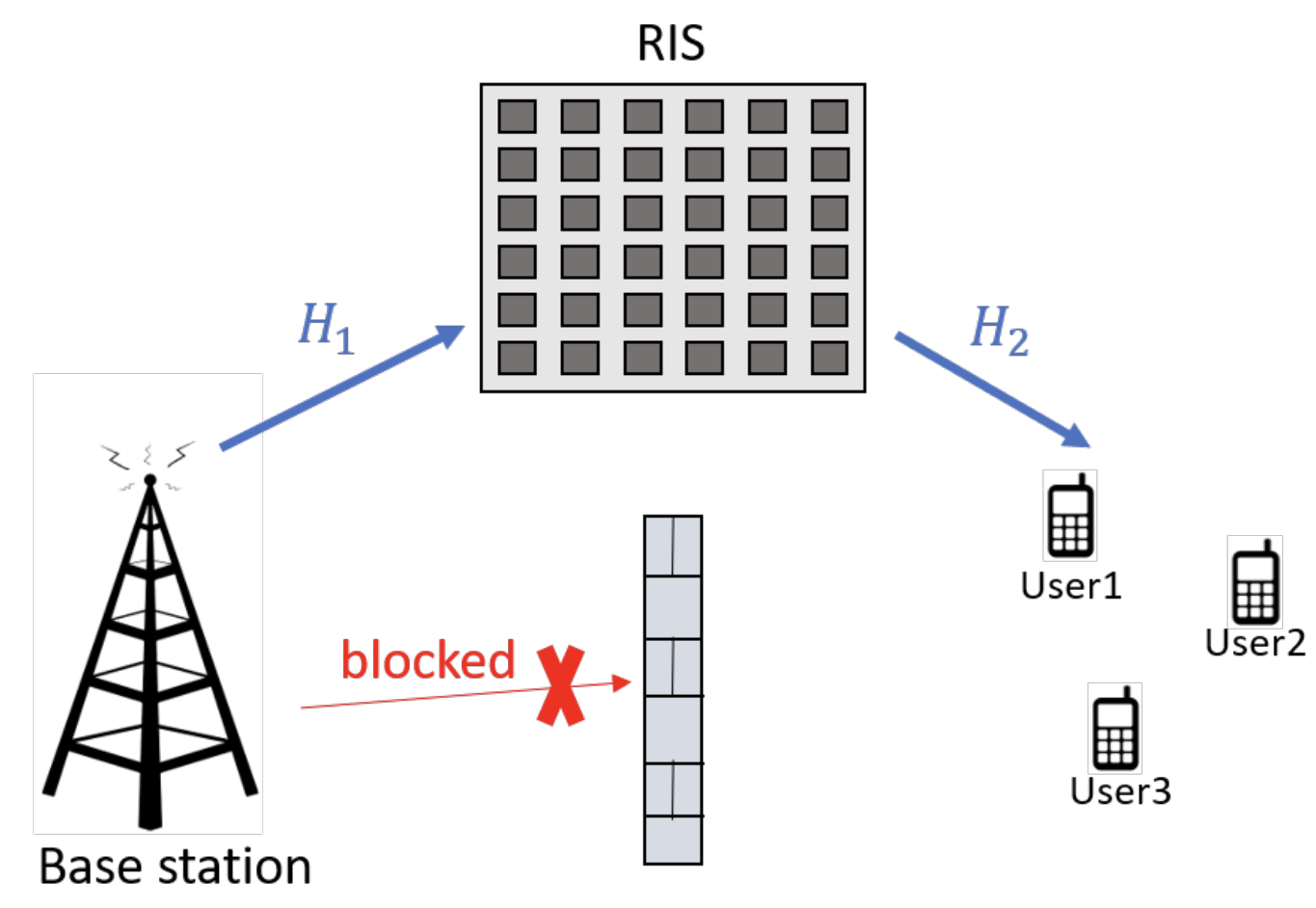}
		\caption[Downlink Communication System Model]{Downlink Communication System Model}
		\label{fig:ris-mimo}
	\end{figure}

	The diagram of the downlink of a RIS-assisted wireless system is shown in Fig. \ref*{fig:ris-mimo}. The BS is assumed to have $M$ antennas. The BS serves $K$ single-antenna users, while the RIS in the system contains $N$ controllable intelligent reflecting elements to provide stable wireless communication from the BS to the users. Furthermore, physical obstacles exist between BS and users, leading to the signal transmitted from the BS not being directly received by the users. Hence, the RIS needs to be deployed between the BS and the users to reduce the effect of physical obstacles and provide reliable paths for the users.
	We assume Rician fading from the BS to the RIS and Rayleigh fading from the RIS to the users. The channels are represented by the channel coefficient matrices ${\mathbf H}_1$ $\in \mathbb{C}^{M \times N}$ and ${\mathbf H}_2 \in \mathbb{C}^{K \times N}$, respectively. When the BS transmits a signal to the $k$th user, the reflected signal from the RIS is received. Therefore, the received signal for the $k$th user is given by:
	\begin{equation}
		\centering
		\label{R1}
		r_k = ({\mathbf h}_{2,k} {\mathbf \Phi} {\mathbf H}_1) {\mathbf W}x + n_k,
	\end{equation}
	
	\noindent where $r_k$ denotes the received signal for the $k$th user and ${\mathbf h}_{2,k} \in \mathbb{C}^{1 \times N}$ the $k$th row of ${\mathbf H}_2$ representing the $k$th user. $\Phi$ denotes RIS reflecting matrix, ${\mathbf W} \in \mathbb{C}^{M \times K}$ is the active digital beamforming matrix for all users. The active digital beamforming vector for the $k$th user is ${\mathbf w}_k \in \mathbb{C}^{M \times 1}$, which is the $k$th column vector of ${\mathbf W}$. Furthermore, $x \in \mathbb{C}^{K \times 1}$ denotes the transmitted signal, which includes the signal for all users and can be expressed as below:
	\begin{equation}
		\centering
		x = \sum_{k = 1}^{K}{\mathbf w}_ks_k, 
		\label{user}
	\end{equation}
	
	\noindent where $s_k$ is the transmitted symbol for the $k$th users. Assume that $s_k$ is generated by independent and identically distributed (i.i.d) random variables with zero mean and unit variance and that it can also be expressed as $s_k \sim \mathcal{CN}(0, 1 )$. We further denote by $n_k \sim \mathcal{CN}(0, \sigma_k^2)$ additive white Gaussian noise (AWGN) for the $k$th user.
	
	In addition, due to the fixed position of the BS and the RIS, the channel ${\mathbf H}_1$ has the component of a line-of-sight (LoS) path. Therefore, channel ${\mathbf H}_1$ can be formulated with path loss term as:
	\begin{equation}
		\begin{multlined}
		\label{H1pathloss}
		{\mathbf H}_1 = \mathcal{L}^{\rm BS-RIS}(d^{\rm BS-RIS}, f_c) \cdot \\
		\left(\sqrt{\frac{K_f}{K_f + 1}}\mathbf{L}_{\rm BS,RIS} + \sqrt{\frac{1}{K_f + 1}}\mathbf{G}_{\rm BS,RIS}\right),  
		\end{multlined}
	\end{equation}
	
	\noindent where the path loss function $\mathcal{L}^{\rm BS-RIS}(\cdot)$ follows the urban propagation model in the 3rd Generation Partnership Project (3GPP) specification TR 36.873\cite{3DPP}. $d^{\rm BS-RIS}$ denotes the 3D distance between the BS and the RIS, $f_c$ denotes carrier frequency. The path loss function $\mathcal{L}^{\rm BS-RIS}(\cdot)$ can be expressed as:
        \begin{equation}
		\centering
		\label{LOSS3GPP1}
        \begin{multlined}
		\mathcal{L}^{\rm BS-RIS}(d^{\rm BS-RIS}, f_c) = \\ 22.0\cdot \log d^{\rm BS-RIS} + 28.0 + \log f_c
        \end{multlined}
	\end{equation}
	\noindent $K_f$ is the Rician factor and $\mathbf{G}_{\rm BS,RIS} \sim \mathcal{CN}(0, 1)$ is the NLoS component.
	Furthermore, the LoS component $\mathbf{L}_{\rm BS,RIS}$ can be expressed as follows:
	\begin{equation}
		\centering
		\label{LOS}
		\mathbf{L}_{BS,RIS} = \mathbf{a}_{RIS}(\phi^A, \psi^A)\mathbf{a}_{BS}(\phi^D)^H,  
	\end{equation}
	
	\noindent where $\mathbf{a}_{RIS} \in \mathbb{C}^{N \times 1}$, $\mathbf{a}_{BS} \in \mathbb{C}^{M \times 1}$, $\phi^{A}$ and $\phi^{D} \in [0, 2\pi)$, and $\psi^A \in \left[ \frac{\pi}{-2}, \frac{\pi}{2}\right)$ represent the steering vector at the RIS and the steering vector at the BS, the elevation angle-of-arrival (AoA)/ angle-of-departure (AoD), and the azimuth. The LoS component is stable due to the fixed positions of the BS and the RIS. We assume that the BS has linear array antennas and the RIS has a uniform planar array controllable reflecting elements with $N_x$ denoting the number of each row and $N_y$ denoting the number in each column. According to\cite{zhang2020capacity}, the equations of linear array antennas and a uniform array of controllable reflecting elements can be expressed as follows:
	\begin{align}
	\begin{split}\label{RIS2}
		{}&\mathbf{a}_{RIS}[n] = e^{j2\pi(n - 1)d_c(\lfloor \frac{n}{N_x} \rfloor \sin(\phi^A)\sin(\psi^A)} \cdot \\
		&e^{(n - \lfloor \frac{n}{N_x} \rfloor N_x \sin(\phi^A)\cos(\psi^A))/\lambda}, 
	\end{split}\\
		{}&\mathbf{a}_{BS}[m] = e^{j2\pi(m - 1)d_a\sin(\phi^D)/\lambda}  \label{RIS1}
	\end{align}
	
	\noindent where $d_a$ and $d_c$ represent the antenna space and reflecting elements space, $\lambda$ denotes the wavelength.

	Due to the consideration of the users moving in the environment, the channel coefficient matrix ${\mathbf H}_2$ is assumed as a Non-Light-of-Sight (NLoS) channel following the Rayleigh fading model. ${\mathbf H}_2$ with path loss can be expressed as:
	\begin{equation}
		\centering
		{\mathbf H}_2 = \sqrt{\mathcal{L}^{\rm RIS-User}(d^{\rm RIS-User}, f_c)}\cdot \mathbf{G}_{RIS-User} 
		\label{NLOS},
	\end{equation}
	
	\noindent where $\mathbf{G}_{RIS-User} \sim \mathcal{CN}(0, 1)$ is the scatter path. The path loss function $\mathcal{L}^{\rm RIS-User}(\cdot)$ also follows the urban propagation model in \cite{3DPP}. $d^{\rm RIS-User}$ denotes the 3D distance between the RIS and the user. The path loss function $\mathcal{L}^{\rm RIS-User}(\cdot)$ can be expressed as:
	\begin{equation}
		\centering
		\label{LOSS3GPP2}
        \begin{multlined}
		\mathcal{L}^{\rm RIS-User}(d^{\rm RIS-User}, f_c) = 36.7\cdot \log d^{\rm RIS-User} \\ + 22.7 + 26\cdot \log f_c - 0.3(z_r - 1.5), 
        \end{multlined}
	\end{equation}
	\noindent where $z_r$ represents the height of the RIS.
	
	On the other hand, the RIS reflecting element $\Phi \overset{\Delta}{=} {\rm diag}[\mathtt{\phi}_1, \mathtt{\phi}_2, ..., \mathtt{\phi}_N] \in \mathbb{C}^{N \times N}$ is used to represent the passive phase shift matrix. $\phi_n = e ^{j \theta_n}$ representing the $n$th controllable reflecting element, where $\theta_n \in [0, 2\pi)$ stands for the passive phase shift of the $n$th reflecting element of the RIS. 
	
	The transmitted signal combines all user-transmitted symbols as expressed in \eqref{user}. Hence, \eqref{R1} can be further divided into three parts: desired signal, interference signal, and noise. Equation \eqref{R1} can be expressed as:
	\begin{equation}
		\centering
		\label{rFinal}
		r_k = ({\mathbf h}_{2,k} {\mathbf \Phi}  {\mathbf H}_1)w_ks_k + ({\mathbf h}_{2,k} {\mathbf \Phi} {\mathbf H}_1)\sum_{i = 1, i \neq k}^{K}w_is_i + n_k, 	
	\end{equation}
	\noindent where the first term is the desired signal as it contains the transmitted symbol $s_k$ for the $k$th user, the second term is the interference signal, and the third term is noise.

\subsection{Problem Formulation}
	
According to \eqref{rFinal}, the instantaneous signal-to-interference-plus-noise ratio
(SINR) of the $k$th user is
\begin{equation}
	\centering
	\label{SINR}
	{\rm SINR}_k = \frac{\abs{({\mathbf h}_k {\mathbf \Phi} {\mathbf H}_1) \cdot {\mathbf w}_k}^2}{\sum_{i = 1, i \neq k}^K\abs{({\mathbf h}_k {\mathbf \Phi} {\mathbf H}_1)\cdot {\mathbf w}_i}^2 + \sigma_k^2}
\end{equation}

	\noindent Therefore, the achievable rate can be obtained as follows:
\begin{equation}
	\centering
	\label{SINRfinal}
	R_k = \log_2(1 + {\rm SINR}_k).
\end{equation}

AR models are adopted to model the time-varying channels:
	\begin{equation}
		\centering
		{\mathbf H}_1(t + 1) = \rho {\mathbf H}_1(t) + \sqrt{1 - \rho^2}\cdot\widehat{{\mathbf H}_1}(t + 1), t \in \mathbb{N}, 
		\label{CSI1}
	\end{equation}
	\begin{equation}
		\centering
		\label{CSI2}
		{\mathbf H}_2(t + 1) = \rho {\mathbf H}_2(t) + \sqrt{1 - \rho^2}\cdot\widehat{{\mathbf H}_2}(t + 1), t \in \mathbb{N}, 
	\end{equation}
	
	\noindent where $\rho$ is defined as the temporal evolution coefficient with $0 < \rho < 1$ and $t$ is the coherence time index. $\widehat{{\mathbf H}_1}(t + 1)$ and $\widehat{{\mathbf H}_2}(t + 1)$ are both i.i.d random variables with the same form as ${\mathbf H}_1$ and ${\mathbf H}_2$, respectively.

	The goal is to optimize the active beamforming ${\mathbf W}$ jointly and the passive phase shift ${\mathbf \Phi}$ to maximize the average sum rate of the RIS-assisted wireless communication system. Thus, the objective function can be formulated as follows:
 
\begin{equation} \label{profor}
\begin{aligned}
\argmax_{\forall t , {\mathbf W}_t, {\mathbf \Phi}_t} \quad & \lim_{\mathcal{T} \to \infty}\frac{\sum_{t = 1}^{\mathcal{T}}\sum_{k = 1}^{K}\log_2(1 + {\rm SINR}_{k,t})}{\mathcal{T}}\\
\textrm{s.t.} \quad & {\rm tr}\{{\mathbf W}_t^H{\mathbf W}_t\}\leq P_{\max} , \forall t \in \mathbb{N};\\
& \abs{\mathbf{\phi}_{t,n}} = 1, \forall n = 1, 2, \dots, N, \forall t \in \mathbb{N} \\
\end{aligned}
\end{equation}

	Two constraints have been considered in our problem formulation. The first constraint ensures that the transmitted power of the BS is lower than $P_{\max}$, which is the maximum transmitted power. The second constraint states that the gain of each reflecting element is set as unit-modulus.
    
    It can be envisioned that one or many RISs will be deployed in future wireless communication systems. Therefore, the study of RIS-assisted systems under time-varying channels is critically important. Note that problem \eqref{profor} is challenging to solve, as the non-stationary environment resulting in the optimization problem varies on a timestep basis. Existing solutions in the literature only consider a similar problem under a time-invariant channel. To tackle this problem, we utilize an MRL-based algorithm to optimize the passive phase shift and active beamforming on a per-timestep basis.

\section{MRL-based Design Solution}
	Continuous Environment Meta Reinforcement Learning (CEMRL) \cite{bing2022meta} is the first MRL algorithm that provides zero-shot adaptation in a time-varying environment. Unlike adaptive MRL, CEMRL works as task inference MRL, which, in our design, can robustly identify occurring CSI in the testing stage and output corresponding actions. This research formulates the wireless communication problem into an MRL Markov Decision Process (MDP) and redesigns CEMRL as our proposed Forced Adaptation Reinforcement Meta-Learning (FARM).
	
\subsection{Markov Decision Process Formalization}	
    Assuming that an agent controls the phase shift of the RIS and the beamformer, the optimization in \eqref{profor} can be transformed into an MDP framework with the channel state information being viewed as the agent's environment. 
	To formalize the optimization problem into an MRL MDP in each timestep, the entire wireless communication system is seen as a task $z$ with the CSI as a state in the state space and the RIS phase shift and active beamforming as an action in the action space. In addition, the achievable rate defined in \eqref{SINRfinal} is being transformed into the reward function. 
	
	The observation state in the defined wireless system consists of the two matrices ${\mathbf H}_1$ and ${\mathbf H}_2$. Furthermore, an action consists of the matrices ${\mathbf W} \in \mathbb{C} ^ {M \times K}$, representing the active beamforming, and ${\mathbf \Phi} \in \mathbb{C} ^ {N \times N}$, representing the passive phase shift matrix. Due to the structure of the adopted neural network, both elements should be in vector forms and split into the real part  $\Re{(\cdot)}$ and the imaginary part $\Im{(\cdot)}$. The state space in timestep $t$ is therefore defined as: 
\begin{equation}
    \centering
	\begin{multlined}
	\label{state}
	\mathcal{S}^t = [\Re{(Vec\{{\mathbf H}_1(t)\})}, \Im{(Vec\{{\mathbf H}_1(t)\})}, \\
	\Re{(Vec\{{\mathbf H}_2(t)\})}, \Im{(Vec\{{\mathbf H}_2(t)\})}],
	\end{multlined}
\end{equation}

\noindent where the dimension of the state space is $2 \times(M \cdot N + K \cdot M)$. We will denote the state as $s^t_i$ for state $i$ at timestep $t$. 
On the other hand, the action space is defined as

\begin{equation}
\centering
	\begin{multlined}
	\label{action}
	\mathcal{A}^t = [\Re{(Vec\{{\mathbf W}^t\})}, \Im{(Vec\{{\mathbf W}^t\})}, \\
	\Re{(Vec\{{\mathbf \Phi}^t\})}, \Im{(Vec\{{\mathbf \Phi}^t\})}],
	\end{multlined}
\end{equation}

\noindent where the dimension of the action space is $2 \times (M \cdot K + N)$. We will denote the action as $a^t_i$ for action $i$ at timestep $t$. 

	Based on \eqref{profor}, the system sum rate is utilized as the reward function. At each training or testing timestep $t$, it can be expressed as:

\begin{equation}
	\centering
	\label{reward}
	\mathcal{R}^t = \sum_{k = 1}^{K}{log_2(1 + R_{k, t})},
\end{equation}

\noindent where $R_{k,t}$ is the spectral efficiency at timestep $t$ following \eqref{SINRfinal}.

\subsection{Architecture overview}

\begin{figure}
	\centering
	\includegraphics[width=1\columnwidth]{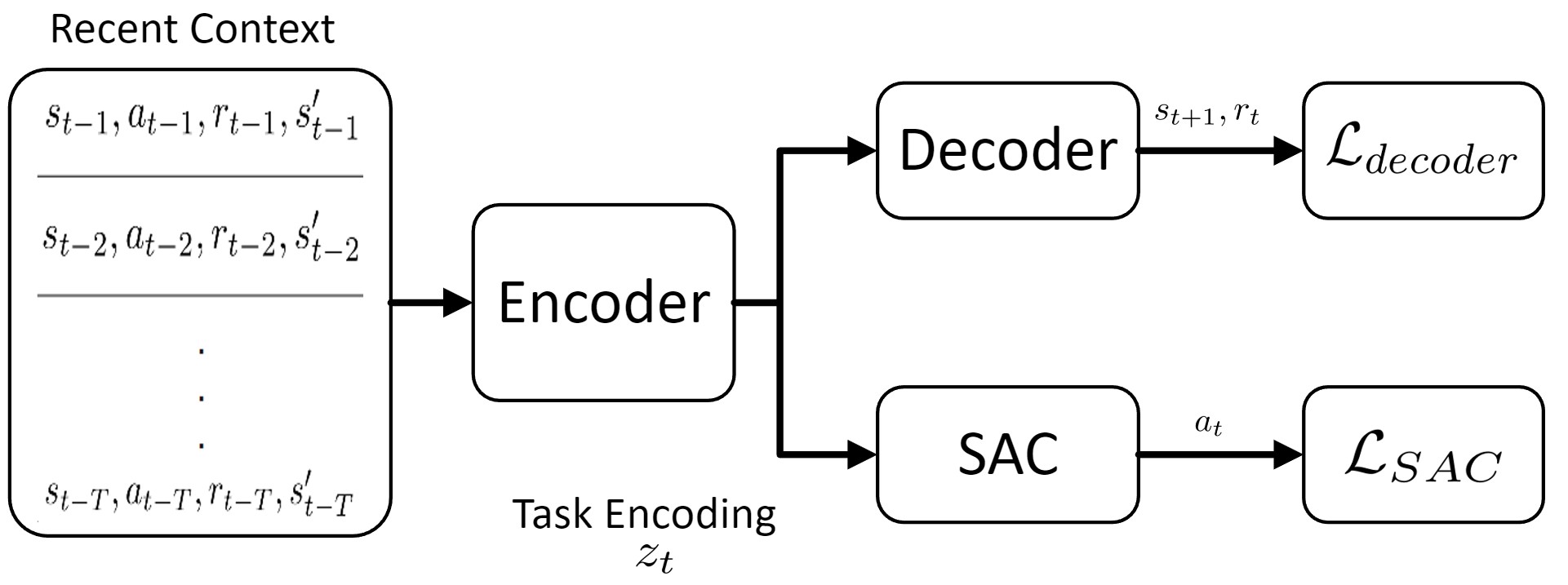}
	\caption[Architecture of Proposed FARM Algorithm]{Architecture of Proposed FARM Algorithm}
	\label{fig:cemrlarchitecture}
\end{figure}

	We adopted the CEMRL architecture proposed in \cite{bing2022meta}. As shown in Fig. \ref*{fig:cemrlarchitecture}, the architecture consists of an encoder $q_\phi$, a decoder $p_\psi$, and a SAC network. The encoder works as a task embedding system and provides the embedded information $z_t$ to the SAC to learn the goal-conditioned policy. The decoder works as an auxiliary loss to provide an unsupervised training manner for the encoder by trying to reconstruct the MDP. 
 
    However, the original architecture's performance will reduce while solving tasks it did not experience during the training stage. To improve the performance of solving unseen CSIs, we redesigned the architecture by adding a task-mapping network in the encoder network. The task-mapping network stores all the experienced CSIs' task encoding during the testing stage. If the encoder produces unseen encoding during testing, the task-mapping network will map the unseen encoding with the stored encoding by the smallest norm.
	
\subsection{Algorithm Pseudo Code and Training-Testing Procedure}

	The training process starts with collecting training data of CSIs from different timesteps following \eqref{CSI1} and \eqref{CSI2}. After data collection, sets of transitions are fed into the encoder to learn the underlying pattern between each element in each transition. Transitions sampled from the same CSI are encoded and merged to produce task indicator $z_t$. The resulting $z_t$ is provided to the conditioned policy to output an action. Thereafter, the decoder is trained with the encoder to reconstruct the underlying MDP in each CSI. Hence, it can train the conditioned policy with the SAC network.
	
	After training, the resulting network is tested in simulation with a fixed meta-policy to operate in a time-varying environment without doing time-consuming calculations. The testing procedure starts with simulating new CSI generated by the AR model with different random seeds from the training stage. The generated CSI is fed into the proposed FARM model for the agent to match the relations between each element with $z_t$. Thereafter, the most optimal action of the MDP, which the decoder constructed from $z_t$, is deployed in the communication system.
 	
	We summarize the pseudo-code in Algorithm 1.
\RestyleAlgo{ruled}

%% This is needed if you want to add comments in
%% your algorithm with \Comment
\SetKwComment{Comment}{/* }{ */}
\begin{algorithm}[h!]
\caption{FARM algorithm for RIS-Assisted Downlink MIMO System Design}\label{alg:1}
\textbf{Input:} Batch of training CSIs $\{T_i\}_{i = 1 \dots T}$ from AR Model\\
\SetKwInOut{Require}{Require}
\Require{Replay buffer $B$}
\For{each training iteration} {
 \ForAll{$T_i$}{
	\For{episode}{
		$y \sim q_\phi(y | c^T), z \sim q_\phi(z | c^T, y), a \sim \pi_\theta(a | s, z)$\;
		Receive next state and reward by action\;
	$c^T$ update with $\tau = (s, a, r, s{'})$\;
	Add $\tau = (s, a, r, s^{'})$ to $B_{episode}$\;
	}
 $B = B \cup B_{episode}$\;
 }
 \Return{$B$}\;
 \ForAll{training steps} {
  Sample $\tau_t \sim B$ and $c^{T_t} \sim B$\;
  \For{$j = 1, \dots J$}{
   $z^{(j)}_t \sim q_\phi(z | c^{T_t}, y = j)$\;
   $p_\psi(s_{t + 1}, r_t | s_t, a_t. z^{(j)}_t)$\;
   $\mathcal{L}^{(j)}_{decoder} = \frac{1}{2} ||\hat{s}_{t + 1} - s_{t + 1}||_2^2 + \frac{1}{2}( \hat{r}_t - r_t)^2$\;
   $\mathcal{L}^{(j)}_{KL_z} = \mathbb{KL}(q_\phi(z, | y = j) || p(z | y = j))$\;
   }
  $\mathcal{L}_{KL_y} = \mathbb{KL}(q_\phi(y | x) || p(y))$\;
  $\mathcal{L}_{ELBO} = \sum_{j = 1}^{J} q_\phi(y = k | c^{T_t})\left[-\mathcal{L}^{(j)}_{decoder} - \alpha_{KL}\mathcal{L}^{(j)}_{KL_z} \right]- \beta_{KL}\mathcal{L}_{KL_y}$\;
  $\psi \gets \psi + \lambda \nabla_\psi \mathcal{L}_{ELBO}(\psi)$\;
  $\phi \gets \phi + \lambda \nabla_\phi \mathcal{L}_{ELBO}(\phi)$\;
 }
 \ForAll{$t$ in $T_{B, max}$}{
 $y_t \sim q_\phi(y | c^{T_t})$\;
 $z_t \sim q_\phi(z | c^{T_t}, y_t)$\;
 Add $(y_t, z_t)$ to transition data in $B: \tau_t= (s_t, a_t, r_t, s^{'}_t, y_t, z_t)$\;
 }
 $\theta^{\pi} \gets \theta^{\pi} - \lambda^\pi \nabla_\theta^\pi J_\pi(\theta^\pi)$\;
 $\theta^{Q} \gets \theta^{Q} - \lambda^Q \nabla_\theta^Q J_Q(\theta^Q)$\;
 $\alpha \gets \alpha - \lambda \nabla_\alpha J_\pi(\alpha)$\;
 }
\end{algorithm}

\section{Simulation Results}
\subsection{Benchmarks and Setting}
	
	In this section, we evaluate the performance of our proposed FARM phase shift and beamforming for a RIS-assisted MU-MIMO wireless communication system. We consider three benchmark schemes:
	\begin{enumerate}
		\item ZF + Random:
		The RIS phase shift matrix is generated randomly with Zero-Forcing (ZF) beamforming. ZF beamforming is a widely used transmit scheme in the literature. This benchmark serves as a lower bound and will be referred to as ZFR in the following. 
		
		\item ZF + Sequential Fractional Programming (SFP):
		SFP is an optimization-based iterative algorithm that uses fractional programming to find the passive phase shift with ZF beamforming. This benchmark is derived from \cite{huang2019reconfigurable}, in which the proposed algorithm only considered the RIS phase shift under time-invariant channel assumption. This benchmark will be referred to as SFP in the following.
		
		\item DDPG:
		The DDPG-based algorithm \cite{jia2021irs} is adopted in this comparison. The authors of the paper considered an Ambient Backscatter Communication (AmBC) system with RIS assistance. The authors then proposed the DDPG-based algorithm to maximize the sum rate. We adapted the core method and applied it to our scenario. 
	\end{enumerate}
	
	Although all the benchmark algorithms, except ZFR, only consider time-invariant channels, they are simulated under both time-varying and time-invariant channels. All the numerical results shown in the following are averaged over 100 channel realizations. Default hyperparameters are summarized in Table \ref*{table:Hypar}.
	
\begin{table}[t]
\caption[HyperParameters]{Default hyperparameters setting of the simulations}
\begin{center}
	\begin{tabular}{|l|c|c|}
		\hline
		Description & Parameter&  Value \\
		\hline
		Number of antennas at Base Station & $M$& 8 \\
		\hline
		Number of RIS elements  & $N$& 32 \\
		\hline
		Number of Users & $K$ & 4 \\
		\hline
		Max transmitted power  & $P_{\max}$ & 10 W \\
		\hline
		Temporal evolution coefficient & $\rho$ & 0.95  \\
		\hline
		Noise power density & $\sigma_k^2$& $-174$ dBm/Hz \\
		\hline
		Carrier frequency & $f_c$ & 5 GHz \\
		\hline
		Rician factor & $K_f$& 3 dB \\
		\hline

	\end{tabular}
	
\end{center}
\label{table:Hypar}
\end{table}

\subsection{Simulation Results and Discussion}

\begin{figure}[!bt] \begin{center}
	
		\includegraphics[width=1\columnwidth]{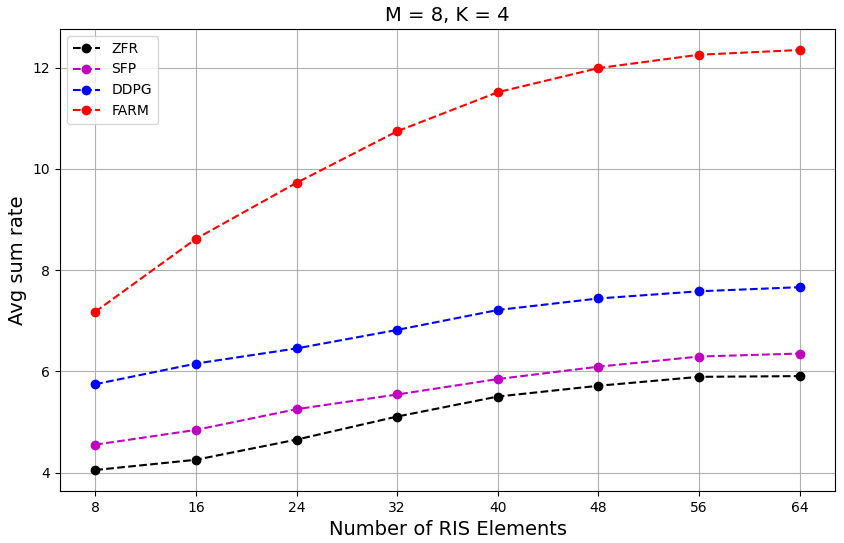}
		\caption[Average Sum Rate Under Different Number of RIS Elements]{Average Sum Rate Under Different Number of RIS Elements}
		\label{fig:RISAvgSumRate}
	\end{center} \end{figure}

	The average sum rate with different numbers of RIS elements is shown in Fig. \ref*{fig:RISAvgSumRate}. The proposed algorithm achieves the highest average sum rate compared with all other benchmarks. This can be explained as follows: the first benchmark only serves as a lower bound, which randomly produces RIS phase shifts without any logical reason. The second benchmark, SFP, optimizes the passive phase shift and power allocation in an iterative way, which results in the algorithm easily producing suboptimal or local optimal output. The last benchmark, DDPG, a DRL-based algorithm, only works as an instant optimizer and cannot work very well with time-varying CSI, even if it is similar to the CSI when the optimization problem is solved. Further, the average sum rate increases with the number of RIS elements, which means there are more signal paths for the mmWave to deliver, resulting in an SINR increase.
	
\begin{figure}[!bt]
		\centering
		\includegraphics[width=1\linewidth]{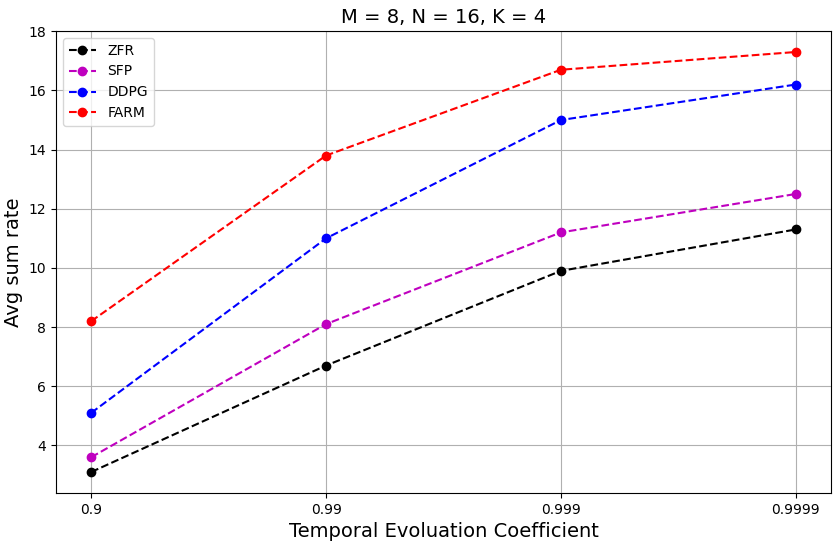}
		\caption[Average Sum Rate under Different Temporal Evolution Coefficient]{Average Sum Rate under Different Temporal Evolution Coefficient}
		\label{fig:tempEvoCoe}
	\end{figure}

	To further illustrate the superiority of our proposed algorithm in a time-varying environment, the temporal evolution coefficient is adjusted with other parameters fixed as the default setting in Table \ref*{table:Hypar}. Fig. \ref*{fig:tempEvoCoe} compares our algorithm with all three benchmarks. The difference in the performance between the proposed FARM algorithm and DDPG benchmarks increases as the value of temporal evolution coefficient $\rho$ decreases. This can be explained by the lower temporal evolution coefficient $\rho$, making the environment more likely to act as a time-varying channel. 
	
\section{Conclusion}

	This paper studied active beamforming and passive phase shift in the downlink RIS-assisted wireless communication system under time-varying channels. The instant optimization problem is viewed as a DRL problem formulated as an MDP. Then, the optimization problem under time-varying channels is considered an MRL problem. FARM, an MRL-based algorithm, is proposed to handle different CSIs and provide corresponding solutions. The proposed algorithm outperforms all the benchmarks under time-invariant and -varying channels. To be precise, it offers 60\% more average sum rate than current state-of-the-art algorithms. It is also applicable under different temporal evolution coefficients, which demonstrates that the proposed method is practical in a time-varying environment. 
	
\section*{Acknowledgment}
	This work was partly supported by MOST 111-2221-E-A49-156-MY3 of Taiwan. The work of Luisa Schuhmacher has received funding from the European Union's Horizon 2020 Marie Skłodowska Curie Innovative Training Network Greenedge (GA. No. 953775).
	
\bibliographystyle{IEEEtran}
\bibliography{citation}

\end{document}